\documentclass[a4paper,aps,prl,showpacs,superscriptaddress,floatfix,nofootinbib,twocolumn,bibtex]{revtex4}

\usepackage{bbold}
\usepackage{bbm}
\usepackage[pdftex]{graphicx}
\usepackage{latexsym,amsmath,verbatim}
\usepackage{color}
\usepackage{rotating}
\usepackage{verbatim}
\usepackage{multirow}
\usepackage[english]{babel}
\usepackage{comment}

\begin{document}

\title{Underestimating extreme events in power-law behavior due to
    machine-dependent cutoffs}

\author{Filippo Radicchi}
\affiliation{Center for Complex Networks and Systems Research, School of Informatics and Computing, Indiana University, Bloomington, USA}
\email{filiradi@indiana.edu.}

\date{\today}

\begin{abstract}
Power-law distributions are typical
macroscopic features occurring in almost all 
complex systems observable in nature, so that
researchers have often the necessity, in quantitative analyses, 
to generate random 
synthetic variates obeying power-law
distributions. The task is usually 
performed through standard methods that 
map uniform random variates into the desired probability space.
Whereas all these algorithms are theoretically solid, in this
paper we show that 
they are subjected to severe machine-dependent
limitations. As results, two dramatic consequences arise: (i)
the sampling in the tail of the distribution is not random but deterministic;
(ii) the moments of the 
sample distribution, that are theoretically expected to diverge as functions 
of the sample sizes, converge instead to finite values.
We provide quantitative indications 
for the range of distribution parameters
that can be safely handled by standard libraries used in computational
analyses. Whereas our findings indicate 
possible reinterpretations of numerical
results obtained through flawed sampling
methodologies, they also open the door for the search 
of a concrete solution to this central issue shared 
by all quantitative sciences
dealing with complexity.
\end{abstract}


\maketitle

If the probability $P(x)$ to observe 
a particular value $x$
of some quantity is inversely proportional to the $\lambda$-th power
of that value, i.e., $P(x) \sim x^{-\lambda}$, 
the quantity under observation is said
to follow a power-law probability density function (pdf)
or distribution. The first 
empirical observation of a power-law 
pdf dates back to the analysis by Pareto at the end
of the $1800$s about the distribution of income
and wealth among the population of Italy~\cite{Moore01111897}.
Not so much after, Zipf noted that the frequency of any word 
in natural languages is inversely proportional to its rank in the 
frequency table~\cite{zipf_1949}. 
Historically speaking, 
the contributions by Pareto and Zipf
represented only the beginning of 
a long series of empirical discoveries of power-law
distributions in natural and man-made
systems. Thanks to the possibility offered by modern 
computational technologies
to retrieve, store and analyze 
large-scale sets of data, the number of such empirical evidences
has drastically increased in the 
last $10-15$ years, and is continuously growing.
Power-law pdfs have been observed by researchers 
in almost all scientific disciplines, 
including physics~\cite{Stanley1987, mandelbrot, barabasi1999emergence}, 
biology~\cite{viswanathan1996levy}, earth and planetary sciences~\cite{gutenberg1944frequency, boffetta1999power}, 
economics and finance~\cite{mantegna1995scaling, reed2001pareto, gabaix2008power}, 
computer science~\cite{faloutsos1999power, mitzenmacher2004brief}, 
and social science~\cite{deSollaPrice1965, brockmann2006scaling}, just to 
mention a few of them. For a recent review of power laws in empirical 
data see~\cite{clauset2009power}.
Power-law distributions 
generally emerge in self-organized, critical,
multiscale, and collective phenomena, and their ubiquity is
interpreted as the consequence of the intrinsic complexity
that drives the dynamics and organization of natural systems.

Computers play a fundamental role not only in the analysis
of data, but they are also often used to run numerical simulations
with the aim of understanding and 
possibly predicting the behavior of complex
systems. Since power-law distributions are the emblematic features
of complexity, computer simulations and analyses 
often rely on the 
construction of sequences of synthetic power-law variates, 
i.e., random numbers extracted from power-law distributions.
Whereas the presence and consequences of
serious numerical errors have been 
already studied in context of record 
statistics~\cite{PhysRevLett.109.164102, PhysRevLett.110.180602},
so far none has questioned the effectiveness of the
algorithms developed for the generation
of power-law random variates. In this paper we show that all
methods currently adopted for this purpose
are instead subjected to severe machine-dependent limitations.

For simplicity, 
we will focus on 
the case of continuous random variates 
obeying the power-law pdf
\begin{equation}
P(x)  = \frac{1-\lambda}{x_M^{1-\lambda} - x_m^{1-\lambda}}\, x^{-\lambda} \; ,
\label{eq:power}
\end{equation}
if $x_m \leq x \leq x_M$, and $P(x) = 0$, otherwise.
$x_M \geq x_m \geq 1$ respectively indicate
the upper and lower bounds of the support
of $P(x)$, whereas $\lambda > 1$
is the exponent of the power-law distribution.
The $k$-th moment of the pdf
of Eq.~(\ref{eq:power}) is given by
\begin{equation}
\langle x^k \rangle  = \frac{1-\lambda}{k+1-\lambda} \, 
\frac{x_M^{k + 1-\lambda} - x_m^{k + 1-\lambda}}{x_M^{1-\lambda} - x_m^{1-\lambda}} \; ,
\label{eq:moments}
\end{equation}
valid for $k +1 \neq \lambda$. For  $k +1 = \lambda$, we have instead
$\langle x^k \rangle  = \frac{\left(1-\lambda\right) \left(\log{x_M} - \log{x_m}\right)}
{x_M^{1-\lambda} - x_m^{1-\lambda}}$.
The previous equation mathematically
formalizes a fundamental 
property of power-law distributions:
all $k$-th moments, with $k + 1 \geq \lambda$, 
diverge as powers (or logarithms) of 
the upper bound $x_M$.
The divergence of the moments of power-law pdfs is at the
basis of many fundamental theoretical results.
One requirement in the
computer generation of power-law random numbers
is thus to preserve this property.
This task is, however, possible only under very restrictive conditions.
Let us illustrate in details the origin of these numerical
problems.

Suppose we want to create a sequence
of computer-generated random variables
extracted from the pdf of Eq.~(\ref{eq:power}).
Whereas there are many different ways to 
perform this task,
here we will concentrate our attention on the so-called 
inversion method~\cite{devroye:1986, clauset2009power}.
We will come back to this point later, but
we can already anticipate that
the problems arising in the inversion method
are common to all other methods that can be possibly
be used to generate non uniform random variates.
Our choice to focus on this method is not arbitrary, but
motivated by its popularity in numerical analyses~\cite{clauset2009power}. 
The algorithm consists of two steps: (i) extract a random variable $q$ from a uniform 
distribution defined over the interval $(0,1)$;
(ii) compute the random variate $x$
from the desired distribution $P(x)$ by inverting the relation
$\int_{x}^{x_{M}}  dy \, P(y)  = q$.
While the approach is valid for any $P(x)$, 
it certainly results to be of great
applicability in all cases in which the integral on the l.h.s. of the
previous equation
can be expressed explicitly, so that $x$ can be written in terms of $q$.
For power-law distributions , we can write
\begin{equation}
x = \left[  x_{M}^{1-\lambda} - (x_{M}^{1-\lambda} - x_{m}^{1-\lambda}) q \right]^{\frac{1}{1-\lambda}}\; ,
\label{eq:inversion}
\end{equation}
and use  this relation to 
generate power-law distributed random variables $x$
directly from uniformly distributed random variates $q$.
Note that, in typical situations, $x_M \gg x_m$ so that
Eq.~(\ref{eq:inversion}) can be written as
$x \simeq x_m \, q^{1/(1-\lambda)}$.

The inversion method is very elegant
and theoretically solid, but, computationally
speaking, it suffers of a severe weakness: its
effectiveness relies 
on the goodness of the 
generator of random uniform variables. 
Generally, this fundamental role is played
by pseudo-random number generators (PRNG). These are deterministic  
algorithms capable to produce high quality 
uniform variates with very long
periods~\cite{devroye:1986, matsumoto1998mersenne}.
However, the  precision at which 
these numbers are created is finite. 
Good PRNGs available on the market
today have a precision of $B_r\geq 32$ bits, meaning
that the minimal distance between two
random numbers generated by the PRNG is
at maximum $2^{-B_r}$. While this precision
is completely satisfactory for the generation
of uniform random variates, when applied 
to the transformation of Eq.~(\ref{eq:inversion}), it
translates into a series of machine-dependent 
thresholds at which the distribution of the samples 
produced by the machine drastically diverges 
from the theoretical distribution $P(x)$.

\begin{figure}[!htb]
\begin{center}
\includegraphics[width = 0.48\textwidth]{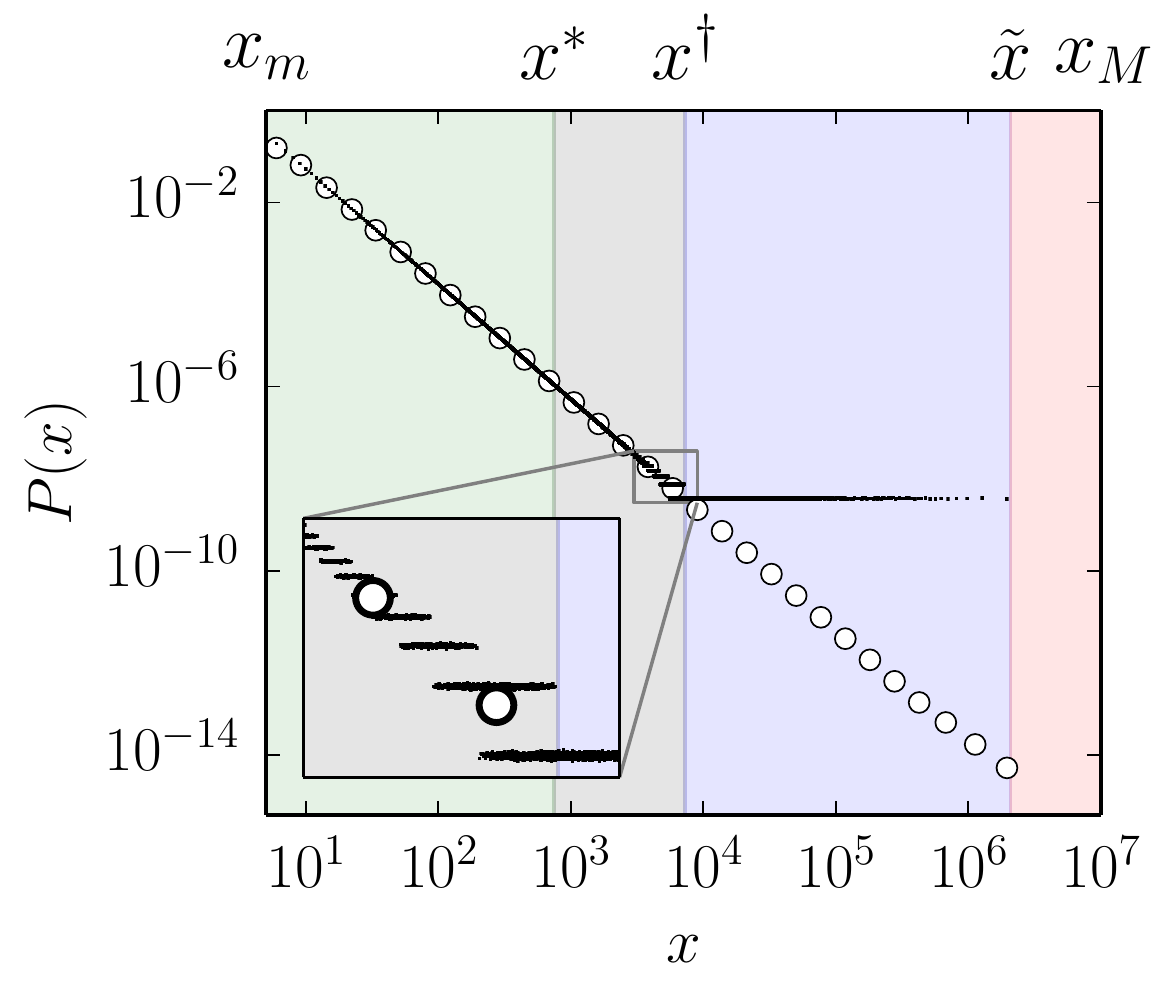}
\end{center}
\caption{Sample distribution obtained from $10^{12}$
random variates extracted from a power-law pdf with 
parameters $x_m=5$, $x_M=10^7$ and $\lambda=2.5$. 
For the generation of the synthetic variates, we used a PRNG 
with $B_r=28$ bits of precision (see Supplemental Material). 
For simplicity of illustration, 
$x$ values have been rounded to the closest integer value, and the 
sample pdf is normalized by the sum of 
these integer numbers. Except for the region 
$x < x^*$ (green shaded area),
the histogram based on linear binning (black squares)
reveals the presence of data discretization as discussed 
in the text (see inset for a zoom of this region).
The final plateau corresponds to a probability
equal to $2^{-B_r}$.
The histogram based on logarithmic binning
hinders the discretization of the data
(white circles). The fact that the exponent measurable
from the histogram based on logarithmic binning 
is equal to $\lambda$ also in the region between $x^\dag$ and
$\tilde{x}$ (blue shaded region) indicates that the 
distance between consecutive
admissible variates in the final plateau 
is growing power-like with exponent $\lambda$.
No points are visible in the region 
$x>\tilde{x}$ (red shaded area).
}
\label{fig:distr}
\end{figure}

The first problem that arises
is the dramatic worsening of the accuracy at which
random variates are extracted from $P(x)$. 
Suppose that we want to be able to
discriminate between two consecutive
random numbers with an accuracy equal to $a$.
This is possible only
if the distance in the probability
space between $x$ and $x+a$ is larger
than $2^{-B_r}$.
This requirement is firstly 
violated at $x=x^*$, with $x^*$ defined by
$P(x^*) - P(x^*+a)  = 2^{-B_r}$.
When $a \ll x^*$, the term of the l.h.s. can be approximated
by the negative derivative of $P(x)$ computed at $x^*$.
For power-law pdfs, we thus have
\begin{equation}
x^* \simeq 2^{\frac{B_r}{\lambda+1}} \, x_m^{\frac{\lambda-1}{\lambda+1}} \, \left[a \lambda (\lambda - 1) \right]^{\frac{1}{\lambda+1}} \; ,
\label{eq:x_star_p}
\end{equation}
valid in the limit $x_M \gg x_m$. For 
values of $x > x^*$, power-law random variates
are not longer extracted with the required accuracy
$a$. Consecutive admissible values differ
by amounts larger than $a$ giving rise to a
unwanted discretization (see Fig.~1).
The most evident effect of the discretization is the presence
of plateaus in the pdf of the samples. Plateaus are defined over overlapping intervals
on the $x$ axis, and sudden jumps occur among
consecutive plateaus.
As $x$ increases, the discretization gets worse
(i.e., plateaus become wider),
until we reach the final plateau
corresponding to a probability equal to $2^{-B_r}$. 
This level is reached for $x=x^\dag$, where $x^\dag$
is defined by
$P(x^\dag)  = 2^{-B_r}$.
For power-law pdfs, this means
\begin{equation}
x^\dag \simeq 2^{\frac{B_r}{\lambda}} x_m^{\frac{\lambda-1}{\lambda}} \left(\lambda - 1\right)^{\frac{1}{\lambda}} \; ,
\label{eq:x_dag_p}
\end{equation}
valid for $x_M \gg x_m$. 
The discretization of the probability and the sample
space is certainly deleterious.
The tail of the true distribution is not
sampled at random, but in a deterministic
fashion. No matter how many variables we extract, 
the sample pdf will never approach the true distribution, 
and particular values of $x$ will be never extracted.
On the other hand, the consequences of this 
discretization on the effective divergence of the moments of the sample 
distribution are not as serious as those produced by
the third threshold that we are going to describe.

Proceeding in the direction of increasing
$x$, an additional and much more severe threshold appears.
Since no uniform numbers lower than $2^{-B_r}$ 
can be extracted, there is no possibility to generate 
random variates from the distribution $P(x)$ larger than 
$\tilde{x}$ defined by
$\int_{\tilde{x}}^{x_{M}}  dy \, P(y)  = 2^{-B_r}$. 
For the special case of power-law pdfs, this
machine-dependent cutoff in the sample
pdf can be estimated as
\begin{equation}
\tilde{x} \simeq  x_m \, 2^{\frac{B_r}{\lambda -1}} \; ,
\label{eq:cutoff_p}
\end{equation} 
again valid for $x_M \gg x_m$.
The presence of a machine-dependent cutoff 
implies that the $k$-th moment of the 
sample distribution behaves as
$\langle x^k \rangle \simeq  \int_{x_m}^{\tilde{x}} dy \, y^k P(y)  + 2^{-B_r} \tilde{x}^k$ .

For power-law pdfs with $k + 1 > \lambda$, 
the $k$-th moment of the sample distribution,
that is supposed to diverge as $x_M$ increases, instead converges to
\begin{equation}
\langle x^k \rangle \simeq x_m^k \, 2^{\frac{B_r (k+1-\lambda)}{\lambda-1}} \, \frac{k}{k+1-\lambda}
\; .
\label{eq:moments_p}
\end{equation}
A similar expression can be also found for $k + 1 = \lambda$.

To illustrate the practical importance of
the machine-dependent limitations described so far,
we present numerical evidences of problems 
induced by the finite precision 
PRNGs in two very common situations.

Fig.~2 illustrates the errors committed in
a typical situation faced in the generation
of random networks with prescribed degree
distributions. These objects are generally 
constructed following the method developed by Molloy 
and Reed~\cite{molloy1995critical}. 
For a network with $N$ nodes, the first
ingredient in the Molloy-Reed recipe consists
in the creation of a degree sequence composed
of random integers numbers extracted from an {\it a priori}
fixed degree distribution, that,
in the case of scale-free networks, consists in 
a list of random variables
extracted from a power-law pdf 
defined over the interval $[x_m, x_M]$,
with $x_M$ generally set equal to $N-1$. 
As it is well known, many fundamental results
for random scale-free networks 
rely on the fact that the second
moment of the degree distribution is
divergent for any $\lambda \leq 3$~\cite{barabasi1999emergence, 
albert2002statistical, dorogovtsev2008critical}. 
Examples include the percolation threshold~\cite{cohen2000resilience},
the epidemic threshold~\cite{pastor2001epidemic}, 
and the consensus time in the voter model~\cite{sood2005voter}.
As Fig.~2 shows, however, the expected divergence
of the moments of the degree distribution is
numerically satisfied only up to values
of $N \simeq \tilde{x}$, as predicted in Eq.~(\ref{eq:cutoff_p}).
Results obtained with different PRNGs 
confirm that the precise value of the cutoffs can be 
machine- and implementation-dependent (see Fig.~S1).

\begin{figure}
\begin{center}
\includegraphics[width = 0.48\textwidth]{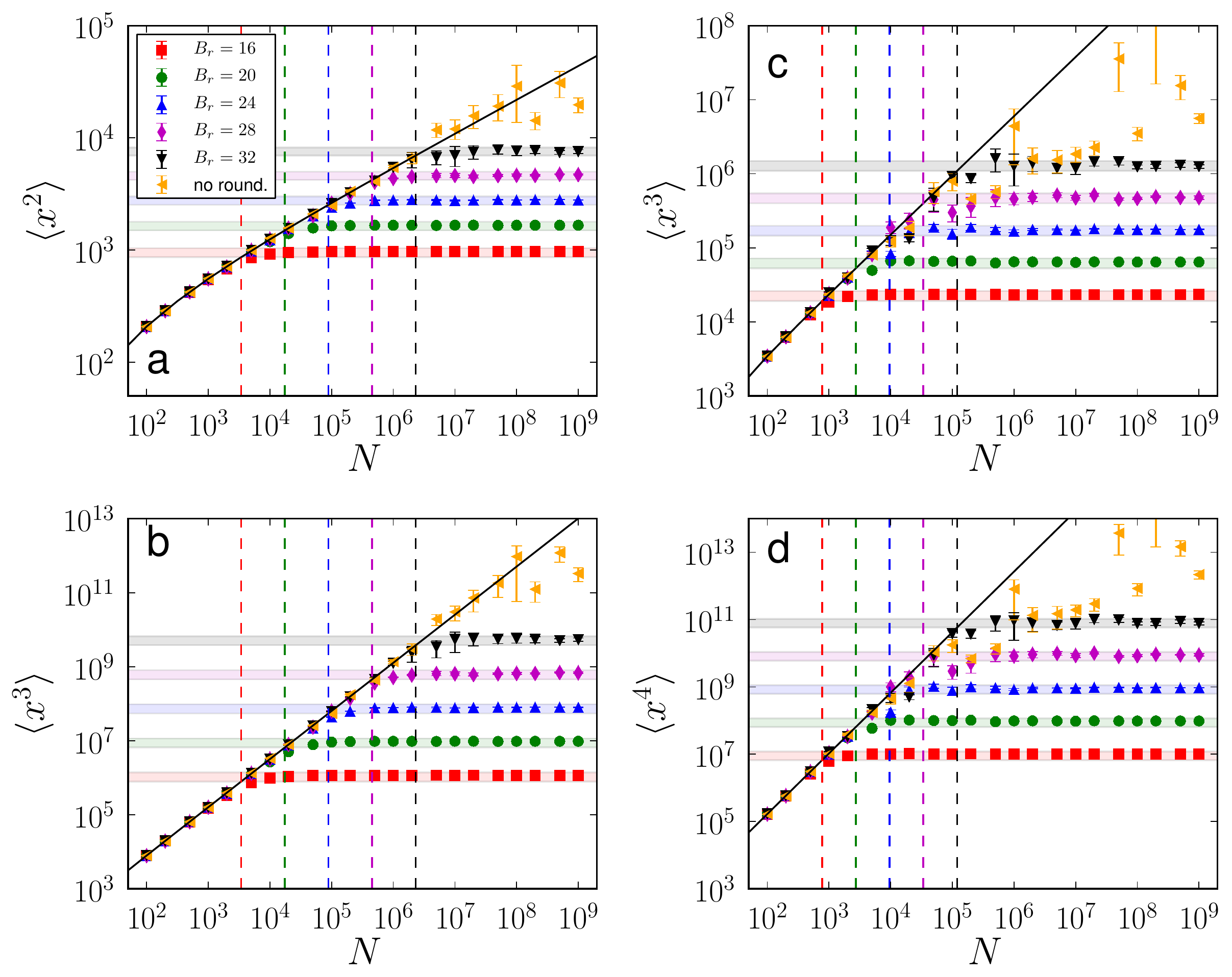}
\end{center}
\caption{Convergence of the
moments in the sample distribution
of $N$ random variates. We fix $x_m=5$ and $x_M=N$. 
The power-law exponent is
$\lambda=2.7$ in panels a and b, and $\lambda=3.2$ 
in panels c and d. In each simulation, we extracted
$N$ random variates and calculated
the moments of the sample distribution. 
Points in the figure 
refer to the average value of the moments of the sample
pdf over at least $100$ independent realizations.
Error bars, when visible,
quantify the standard error associated to this measure.
Black lines, corresponding to
$\int_{x_m}^{x_M} dy \, y^k P(y)$, represents
the expected value of the $k$-th moment
for the theoretical distribution.
Dashed vertical lines stand for 
the machine-dependent cutoffs 
predicted by Eq.~(\ref{eq:cutoff_p}).  
Horizontal shaded areas are delimited
by $\int_{x_m}^{\tilde{x}} dy \, y^k P(y)$ and 
Eq.~(\ref{eq:moments}). Different colors and symbol
types correspond to different values of the number
of bits $B_r$ used in the generation of uniform 
random variates. Orange points are obtained
without rounding the value of the random uniform
variates (see methods for details).
}
\label{fig:conv}
\end{figure}

Our second example regards
the statistics of the sum of power law random variables
$S  = \sum_{i=1}^T x_i$. This is a quantity of interest
in many practical situations. 
For example, the total deformation of rock materials or of the Earth's 
surface due to earthquakes may be modeled as the sum of 
seismic moment tensors~\cite{sotolongo2000levy, kagan2002seismic}, 
the latter obeying the Pareto distribution; 
cumulative economic
losses or casualties due to natural catastrophes are modeled as sums of power-law
distributed variables ~\cite{voit2003statistical}; 
the total pay-off of an insurance company is modeled as the sum of individual
pay-offs, each of which is distributed according to a power law~\cite{kagan1997earthquake}.
Also, the sum of random power-law variates 
plays an important role in superdiffusive processes modeled
by L\'evy flights, i.e., a special class of random walks 
in which the step-lengths obey a 
power-law probability distribution~\cite{shlesinger1993strange}.
Numerical simulations of L\'evy flights
are used in several contexts: for example,
in the study of movements of
animals and humans~\cite{viswanathan1996levy, brockmann2006scaling}
and stochastic models of physical systems~\cite{shlesinger1995levy}.  
Theory predicts two different behaviors
for the sum of power-law random variables:
for values of the power-law exponent $\lambda > 3$, the pdf 
$P(x)$ has finite variance, the central limit theorem
applies, and the distribution $P(S)$ of 
the sum $S$ approaches a normal distribution 
as the number of summands $T$ grows to infinity;
for $\lambda \leq 3$ instead, $P(x)$ has diverging
variance, the generalized central
limit theorem applies, and $P(S)$ approaches a 
so-called $\alpha$-stable distribution
as $T$ grows~\cite{nolan2003stable, zaliapin2005approximating}.
The presence of the upper-bound in the distribution
$P(x)$ induces, however, catastrophic deviations
from the expected behavior (see Fig.~S2). For any value of 
the exponent $\lambda$ there exists
a maximal number of summands $\tilde{T}$ after
which the distribution $P(S)$ starts
to show features typical of a normal distribution:
coefficient of variation, skewness and excess
kurtosis decrease to zero as the
number of summands increases. This essentially
means that if the number of summands
is large enough, the distribution $P(S)$ starts
to become peaked and symmetric around a given value,
as expected for a normal distribution.
On the basis of the value of the machine-dependent
cutoff of Eq.~(\ref{eq:cutoff_p}), we 
should expect $P(S)$ to approach
a normal distribution very slowly~\cite{mantegna1994stochastic}.
Our results, however, suggest that $P(S)$ enters
in a regime of ``normality'' for small
values of the number of summands.
For $B_r=32$ for example,
the distribution starts to approach a normal distribution
at  $\tilde{T} \simeq 10^4$ when $\lambda = 1.5$, and
only $\tilde{T} \simeq 10^3$ for $\lambda=2.5$.

One may argue that the reason of the 
machine-dependent limitations
illustrated so far resides only in the finite precision
of the PNRG, so that
the solution to this problem would be just to use
a PRNG able to produce a number of random bits $B_r$
sufficiently large. Unfortunately, the solution is not as simple. 
When $B_r \geq 32$, as in the case of the orange points of Fig.~2, an additional 
machine-dependent effect may appear: 
the finite precision in the representation
of numbers in our machine that affects the computation
of powers.
This second limitation approximately arises when we reach the so-called 
machine-$\epsilon$ or unit round off of our machine, i.e., 
the bits used in the so-called
mantissa of the floating point number
representation in the machine~\cite{Goldberg:1991:CSK:103162.103163}. 
Under the
hypothesis of having a PRNG able to produce
a number of random bits $B_r > B_\epsilon$, the
previous machine-dependent thresholds 
of Eqs.~(\ref{eq:x_star_p}),
~(\ref{eq:x_dag_p}) and ~(\ref{eq:cutoff_p}) are still
valid by substituting $B_r$ with $B_\epsilon$.
In our implementation, we have 
$B_\epsilon = 52$. 
This is however a limit that can 
be reached  only if the PRNG is
truly able to generate a sufficiently
large number of random bits. The results of our simulations instead
indicate that strong numerical
inaccuracies arise earlier.

To summarize, the current implementations
of the inversion method are not suitable for the
generation of synthetic random variates
obeying power-law distributions.
Problems arise for the violation of two 
basic principles at the basis of the theory of this method:
(i) there exists a perfect uniform random variate generator;
(ii) computers can store and manipulate real numbers.
The same principles are violated for any other
method developed for the computer-generation
of non uniform random variates, so that
we expect to see similar problems also
for other popular algorithms such as the rejection or 
the acceptance-complement methods~\cite{devroye:1986}. 
In this paper, we explicitly considered the case of
continuous random variates, but our results
extend also to discrete variables.
In addition, the same machine-dependent
limitations hold for
other pdfs rather than power-law distributions.
For instance, exponential distributions are subjected to
even stronger discretization effects. On the other hand,
exponential pdfs have finite moments, and
the presence of the machine-dependent cutoff $\tilde{x}$
causes errors not as dramatic as in the case
of power laws or other heavy-tailed distributions.

The practical consequences of
the existence of machine-dependent distortions
in the computer generation of power-law random variates
are diverse. For example, the discretization
of random variables in the tail may be
crucial for the outcome of tests of statistical 
significance of empirical data.
The presence of a finite cutoff is certainly more
dangerous. Finite size scaling analyses, for example, that do
not account for it, are all potentially subjected
to uncontrollable scaling factors.
Even more seriously, in predictive analyses there is the
concrete risk of underestimations
of extreme events. Last but not least, 
since the limitations are machine-dependent neglecting their
presence may cause problems of reproducibility
of numerical experiments on different computers.
The general and counter intuitive
message of our analysis is that increasing the size
of the sample does not improve the statistics, since
in computer simulations fundamental properties of power-law 
distributions are preserved 
only up to relatively small sample sizes. 
With the current methodology, the only way to avoid
all the numerical problems illustrated in this paper 
is to impose a forced cutoff
at $x^*$ as defined in Eq.~(\ref{eq:x_star_p}), reducing
however the support of the pdf to incredibly small ranges. 
Future research must thus work on finding
algorithms that do not suffer of this serious problem.
In addition to more radical solutions based on
the use of arbitrarily long numbers of bits
in the representation of both uniform random numbers and 
powers, we believe that effective approaches
could be based on generative models of power-law 
distributions~\cite{mitzenmacher2004brief}, 
features of dynamical systems~\cite{palatella2012noise},
or rely on the
scale invariance property of powers.

\bibliography{biblio}{}

\newpage

\onecolumngrid

\setcounter{figure}{0}
\setcounter{equation}{0}
\renewcommand{\thesection}{S \arabic{section}}
\renewcommand{\theequation}{S\arabic{equation}}
\renewcommand{\thefigure}{S\arabic{figure}}
\renewcommand{\thetable}{S\arabic{table}}

\section*{Supplemental Material}

\section*{Methods}
To the best of our knowledge, the
most efficient and accurate 
PRNG existing on the market is 
the Mersenne-Twister
algorithm {\tt  mt19937} developed by Matsumoto and 
Nishimura~\cite{matsumoto1998mersenne}. 
In our numerical simulations, 
we used the $64$-bits version of {\tt  mt19937}
coded in {\tt c} by the same inventors
of the algorithm~\cite{mt}, compiled using the version $4.4.7$
of the {\tt gnu c} compiler~\cite{gcc}. Simulations 
were run on $64$-bits machines. To test the dependence of 
the machine-dependent cutoffs on the precision
of the PRNG, we rounded the values produced by {\tt  mt19937} to
the desired number of bits $B_r$.
To further show how results are
strongly machine- and/or implementation-dependent,
in Fig.~S1, we compared the results 
of Fig.~2 with those obtained
using  the default {\tt python} PRNG, 
and the {\tt ran2} PRNG of 
numerical recipes~\cite{Press:2007:NRE:1403886}.

\newpage

\begin{figure*}
\begin{center}
\includegraphics[width = 0.98\textwidth]{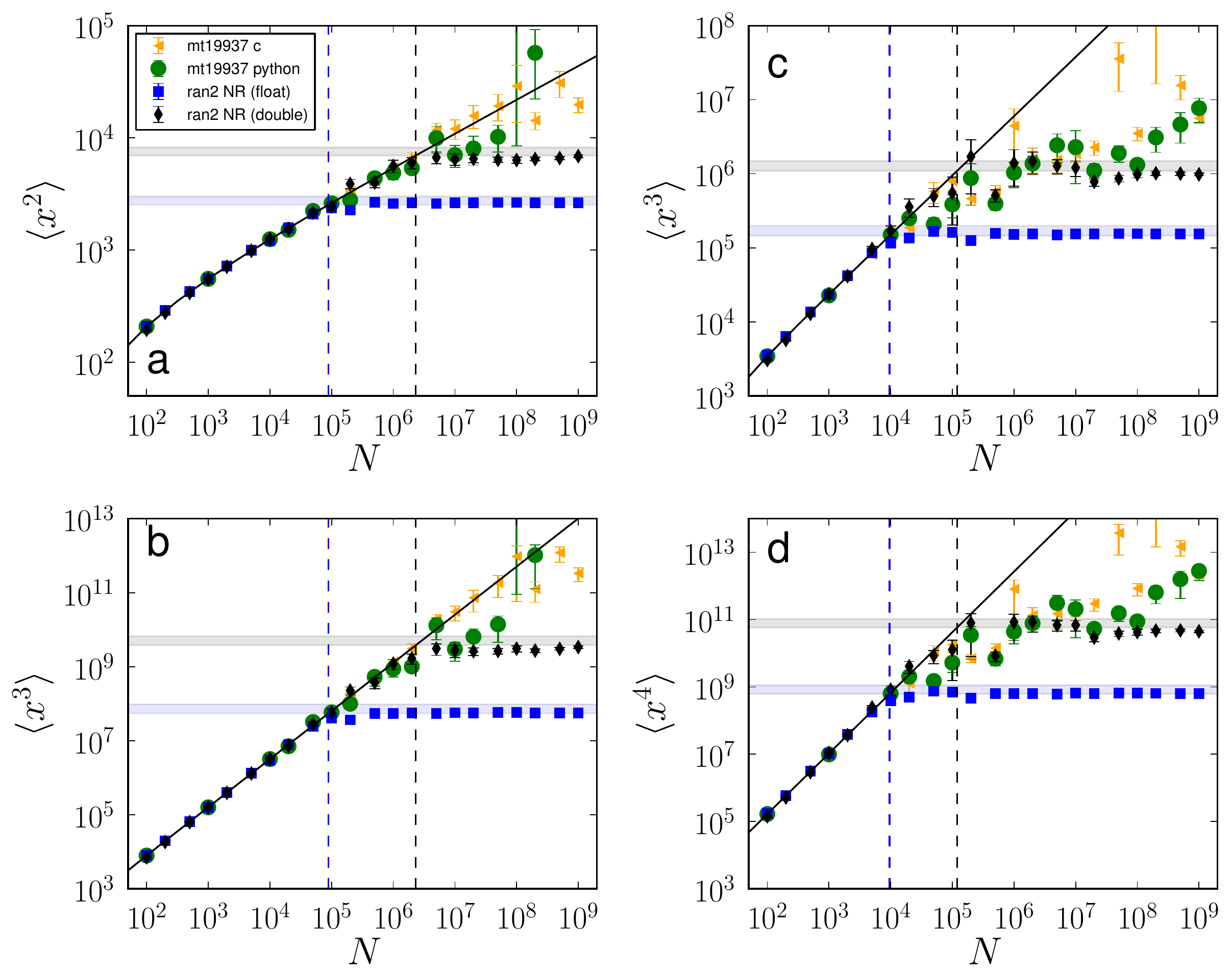}
\end{center}
\caption{Same as Fig.~2 of the main text.
Simulations have been conducted using the 
{\tt c} implementation of {\tt mt19937} (orange triangles),
the 
{\tt python} implementation of {\tt mt19937} (green circles), and
two different
{\tt c} implementations of {\tt ran2} of numerical recipes 
(blue squares and black triangles). The single precision
version of {\tt ran2} (blue squares) generates
random uniform variates with precision of $B_r = 23$ bits, equal
to the number of significant digits of single floating
numbers. The double precision version 
(black triangles) generates random uniform variates with precision 
of $B_r=32$ bits smaller than the one of the double
precision numbers (i.e., $B_\epsilon =52$).
}
\label{fig:S1}
\end{figure*}

\begin{figure*}[!htb]
\begin{center}
\includegraphics[width = 0.96\textwidth]{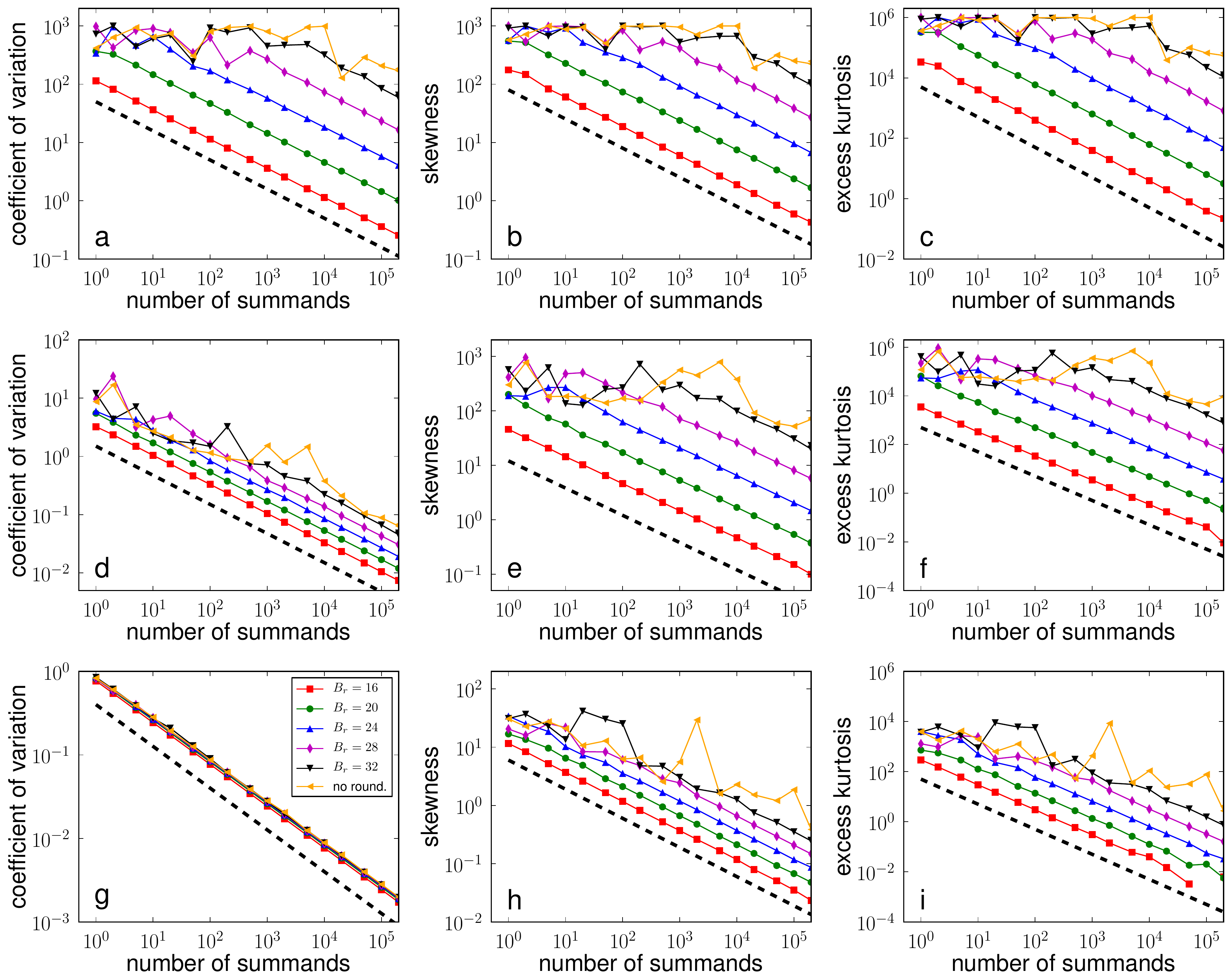}
\end{center}
\caption{Convergence towards the normal distribution. 
We compute the sum $S = \sum_{i=1}^T x_i$
of $T$ i.i.d. random variates extracted from
power-law distributions. The lower and upper
bounds of the support of the power-law distributions
have been respectively set  as $x_m=1$ and
$x_M = \infty$. The various panels show
how the coefficient of variation (a, d and e),
the skewness (b, e, and h), and the excess
kurtosis (c, f and i) of the distribution of $S$
change as functions of the number $T$ of summands.
We consider different power-law exponents:
$\lambda=1.5$ is panels a, b and c, $\lambda=2.5$
in panels d, e and f, and $\lambda = 3.5$ in panels g, h and i.
Depending on the number of bits $B_r$ used in the
generation of the random variables, the plots show
the presence of 
a maximal value $\tilde{T}$ after which the distribution
of $S$ starts to converge to a normal distribution.
The black dashed lines in each panel serve as
guides to the eye to determine the scaling
of the various quantities as functions
of $T$. In panels a, b, d, e, g and h 
the black dashed lines go to zero as the inverse
of the square root of $T$. In panels c, f and i, they
go to zero as $1/T$. Each point is calculated
on the basis of at least $10^5$ independent 
realizations.
}
\label{fig:S2}
\end{figure*}

\end{document}